\documentclass{iau} 
\usepackage{graphicx}

\title[OGLE-ing the Magellanic System: Three-Dimensional Structure] 
{OGLE-ing the Magellanic System: \\ Three-Dimensional Structure}

\author[Anna M. Jacyszyn-Dobrzeniecka \& the OGLE Team]   
{Anna M. Jacyszyn-Dobrzeniecka$^1$ 
\and the OGLE Team$^1$}

\affiliation{$^1$Astronomical Observatory, University of Warsaw \\
Al. Ujazdowskie 4, 00-478 Warszawa, Poland \\
email: {\tt ajacyszyn@astrouw.edu.pl}}

\pubyear{2018}
\volume{344}  
\jname{Dwarf Galaxies: From the Deep Universe to the Present}
\editors{K. McQuinn \& S. Stierwalt}
\begin{document}

\maketitle

\begin{abstract}
We present a three-dimensional structure of the Magellanic System using over 9 000 Classical Cepheids and almost 23 000 RR Lyrae stars from the OGLE Collection of Variable Stars. Given the vast coverage of the OGLE-IV data and very high completeness of the sample, we were able to study the Magellanic System in great details.

We very carefully studied the distribution of both types of pulsators in the Magellanic Bridge area. We show that there is no evident physical connection between the Clouds in RR Lyrae stars distribution. We only see the two extended structures overlapping. There are few classical Cepheids in the Magellanic Bridge area that seem to form a genuine connection between the Clouds. Their on-sky locations match very well young stars and neutral hydrogen density contours. We also present three-dimensional distribution of classical pulsators in both Magellanic Clouds.

\end{abstract}

\firstsection 
\section{Introduction}

The Magellanic Bridge is a direct evidence of LMC-SMC interactions (\cite[Harris 2007]{Harris2007}). Our latest study aims at revealing the three-dimensional distribution of classical pulsators in this structure using Optical Gravitational Lensing Experiment (OGLE) data (\cite[Udalski et al. 2015]{Udalski2015}). Since our previous studies, where we presented the three-dimensional structure of the entire Magellanic System (\cite[Jacyszyn-Dobrzeniecka et al. 2016, 2017]{paperI,paperII}), the OGLE Collection of Variable Stars (OCVS) was updated and a number of newly classified objects was added (\cite[Soszy\'nski et al. 2015, 2016, 2017]{Soszynski2015,Soszynski2016,Soszynski2017}). Thus, we reanalyzed the samples of classical Cepheids (CCs) and RR Lyrae (RRL) stars in the Bridge. Moreover, we also studied the three-dimensional distribution of anomalous Cepheids (ACs).

We also repeat the procedure used by \cite[Belokurov et al. (2017)]{Belokurov2017} to select RRL candidates from Gaia Data Release 1 (DR1, \cite[Gaia Collaboration 2016]{gdr1}) and show that their discovery of a bridge-like structure is based on non-physical sources. Finally, we also show the distribution of classical pulsators from Gaia Data Release 2 (DR2, \cite[Gaia Collaboration 2018]{gdr2}) in the Bridge area and compare it to our results obtained using OCVS.

In Sections~2 and 3 we describe distribution of Cepheids and RRL stars in the Magellanic Bridge. These are the main results from our newest study that is soon to be published (Jacyszyn-Dobrzeniecka et al. in prep.). Sections~4 and 5 highlight main results from our previous studies concerning three-dimensional structure of the LMC and SMC, respectively, using classical pulsators (\cite[Jacyszyn-Dobrzeniecka et al. 2016, 2017]{paperI,paperII}).

\section{Cepheids in the Bridge}

\begin{figure}[htb]
\begin{center}
	\includegraphics[width=.49\textwidth]{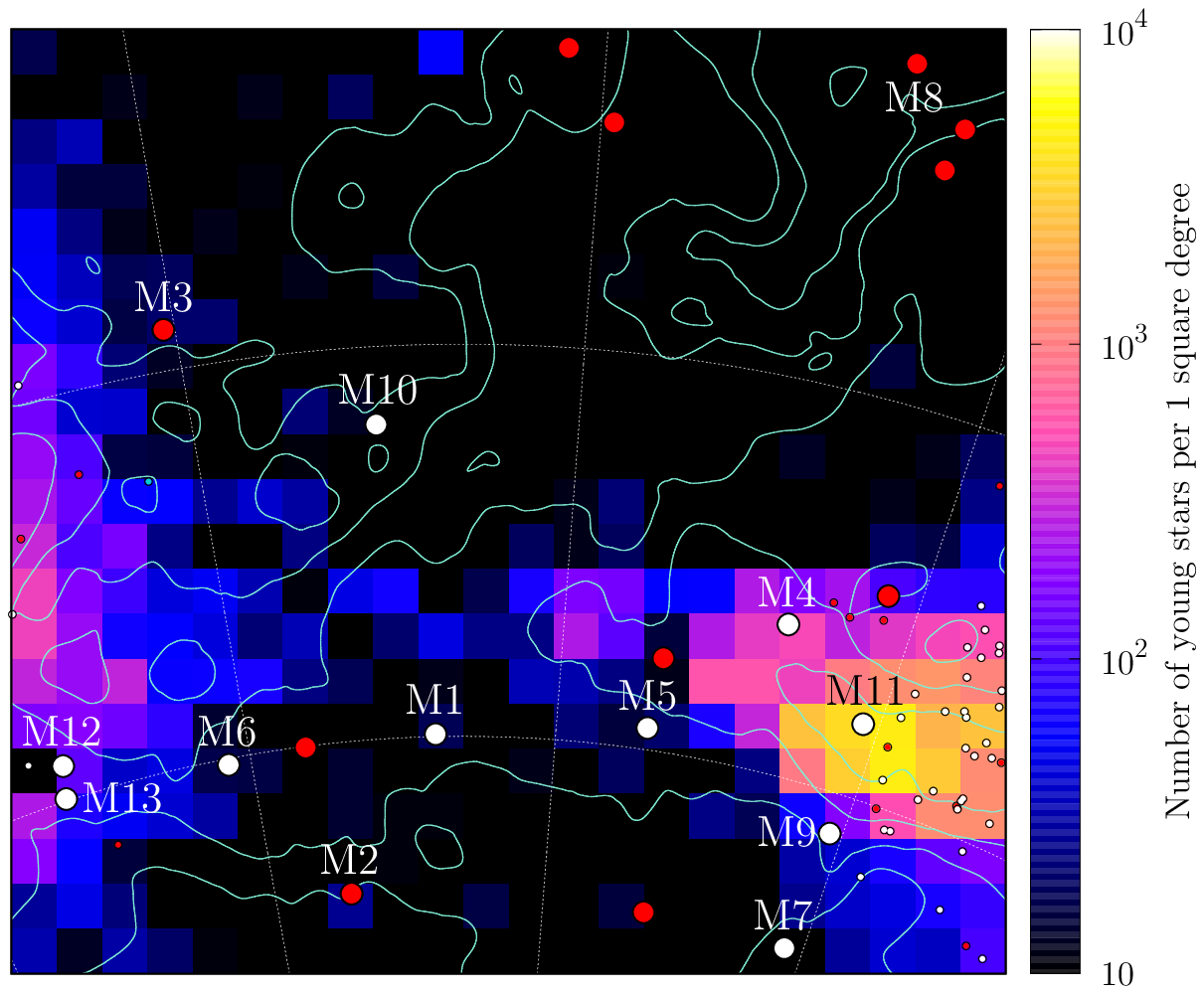} 
	\includegraphics[width=.5\textwidth]{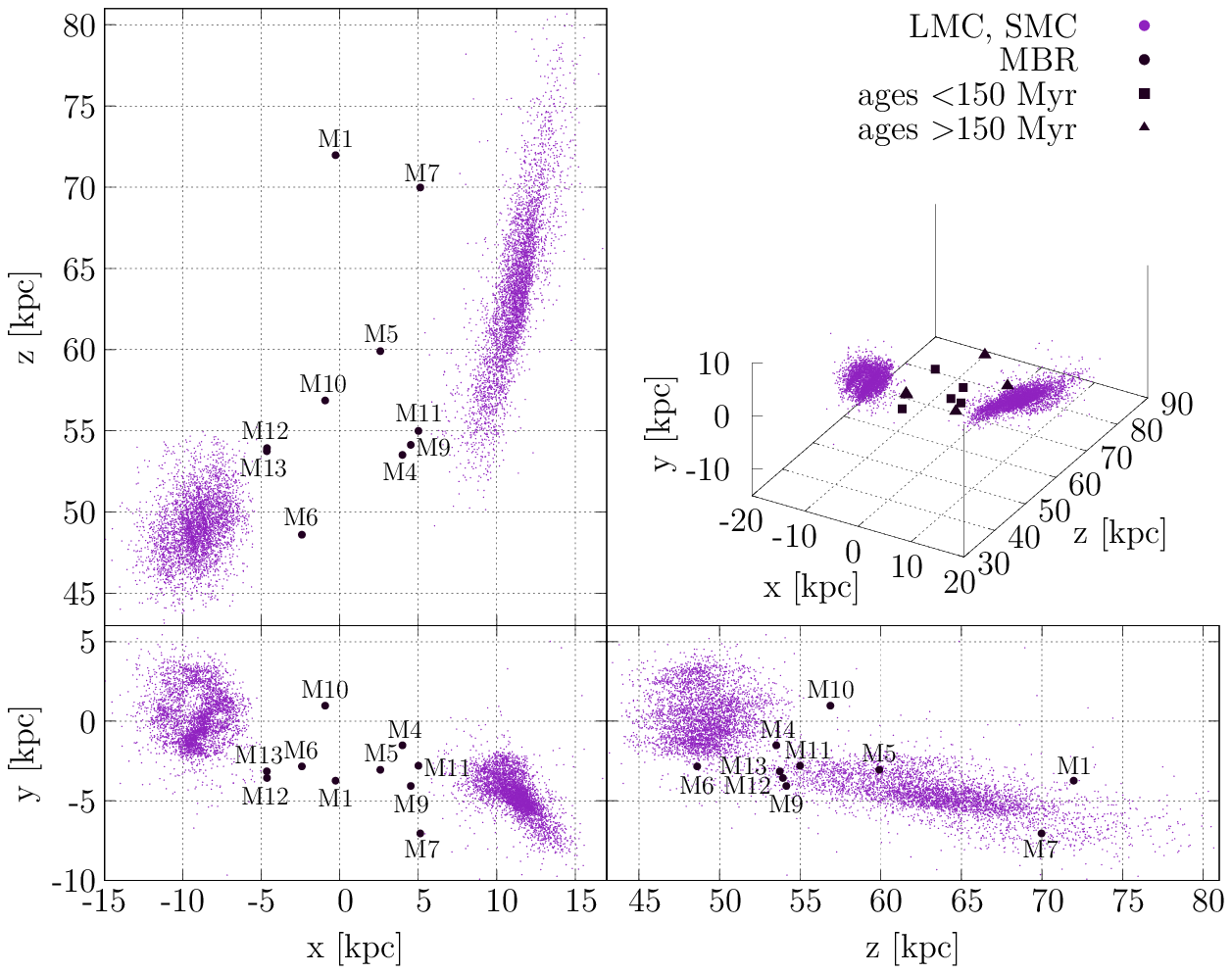} 
	\caption{{\it Left:} OGLE classical (white) and anomalous (red) Cepheids with color-coded young stars column density (\cite[Skowron et al. 2014]{Skowron2014}) and \textsc{Hi} contours (\cite[Kalberla et al. 2005]{Kalberla2005}). {\it Right:} Three-dimensional distribution of classical Cepheids. Figures from Jacyszyn-Dobrzeniecka et al. in prep.}
   \label{mbr-cep}
\end{center}
\end{figure}

Our updated Bridge sample consists of 10 CCs. Their on-sky distribution is very bridge-like and matches the young stars (\cite[Skowron et al. 2014]{Skowron2014}) and \textsc{Hi} (\cite[Kalberla et al. 2005]{Kalberla2005}) highest densities (see left panel of Fig.~\ref{mbr-cep}). In three-dimensions these stars indeed seem to form a connection between the Magellanic Clouds (see right panel of Fig.~\ref{mbr-cep}). However, some of them may be Counter Bridge members or L/SMC outliers.

We also classified 13 ACs as Bridge candidates. Three of them were recently reclassified from CCs. ACs are more spread in both two and three dimensions and do not seem to form a bridge-like connection.

\section{RR Lyrae Stars in the Bridge}

\begin{figure}[htb]
\begin{center}
	\includegraphics[width=.49\textwidth]{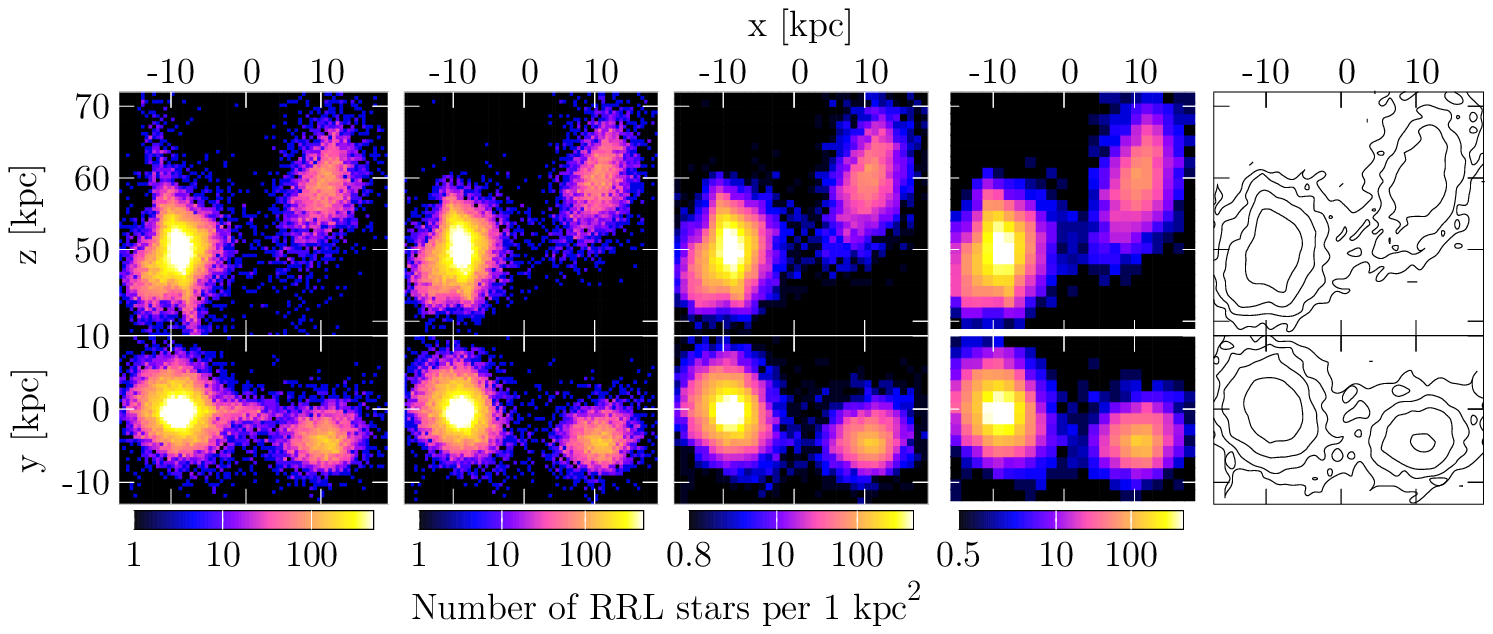} 
	\includegraphics[width=.5\textwidth]{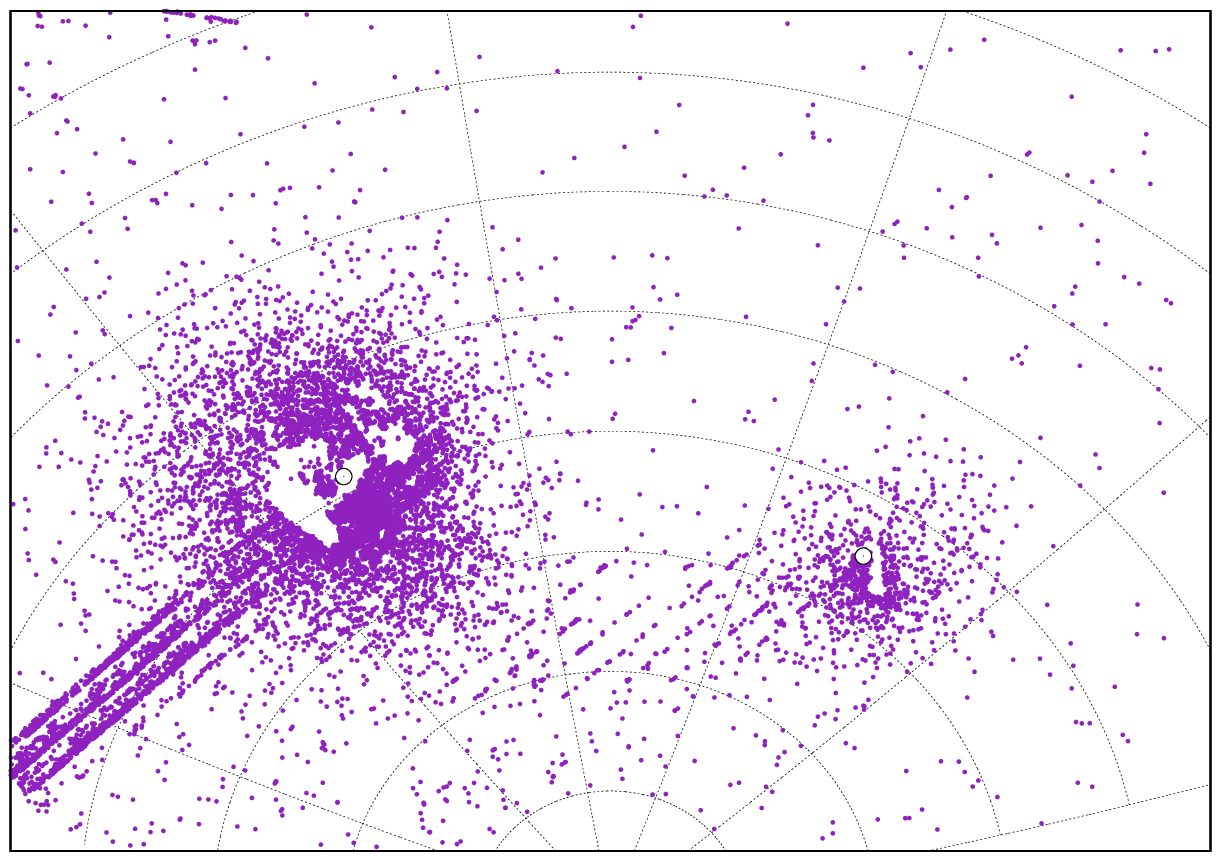} 
	\caption{{\it Left:} OGLE RRL stars in Cartesian coordinates (top and front view). Left panel shows uncleaned sample, other panels -- sample after $3\sigma$-clipping with different bin sizes. {\it Right:} RRL candidates selected from Gaia DR1 using method described in \cite[Belokurov et al. (2017)]{Belokurov2017}. Many non-physical sources are visible forming stripes between the Magellanic Clouds. Figures from Jacyszyn-Dobrzeniecka et al. in prep.}
   \label{mbr-rrl}
\end{center}
\end{figure}

We examine two- and three-dimensional distributions of RRL stars in the Bridge area. Confirming results from \cite[Wagner-Kaiser \& Sarajedini (2017)]{WagnerKaiser2017} and \cite[Jacyszyn-Dobrzeniecka et al. (2017)]{paperII}, we show that we do not see any direct evidence of a bridge-like structure using these tracers. We can only see two extended structures overlapping. We perform a multi-Gaussian fit to our data, proving that there is no additional stellar population located in the Bridge area. Left panel of Fig.~\ref{mbr-rrl} shows that the contours do connect, however, this occurs only on a very low level. RRL stars obtained by Gaia DR2 have a very similar on-sky distribution and no evident bridge-like connection is visible therein.

To test the RRL candidates bridge discovered by \cite[Belokurov et al. (2017)]{Belokurov2017} we apply the procedure presented in their paper to Gaia DR1. Resulting map of the Magellanic System is shown in right panel of Fig.~\ref{mbr-rrl}. Many objects located in stripes between the Magellanic Clouds are non-physical sources caused by cross-match failures in DR1. A cross-match of a selected sample in the central part of the Bridge with OCVS shows that only 15\% of these objects are RRL stars.

\section{Classical Pulsators in the LMC}

CCs in the LMC are situated mainly in a disk-like structure and clumped in substructures. We divided the entire sample into subsamples according to these substructures and carefully analysed each other. The eastern (most prominent and luminous) and western parts of the bar are matching very well in the context of locations and ages of Cepheids. We decided to treat both parts together as a ''new'' bar. The redefined bar shows no offset from the plane of the LMC. The northern arm is also very prominent with an additional smaller arm. Both are located closer to us than the entire sample. The entire LMC sample is rotated towards the SMC.
				
The LMC revealed a very regular structure in RRL stars distribution. We fitted triaxial ellipsoids to our sample. In the LMC we noticed a very prominent, non-physical ''blend-artifact'' that prevented us from analyzing the central parts of this galaxy.

\begin{figure}[htb]
\begin{center}
	\includegraphics[width=.49\textwidth]{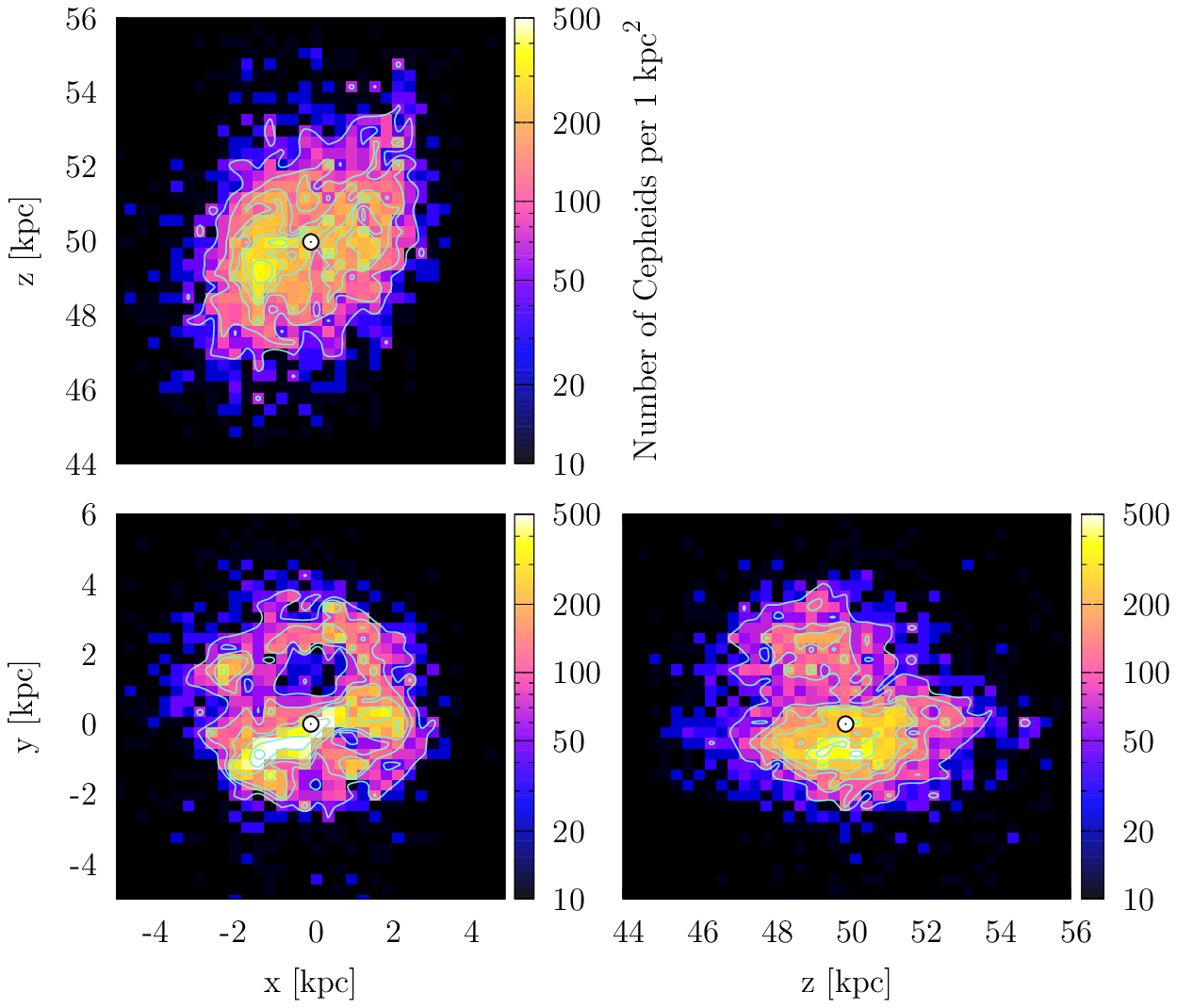} 
	\includegraphics[width=.5\textwidth]{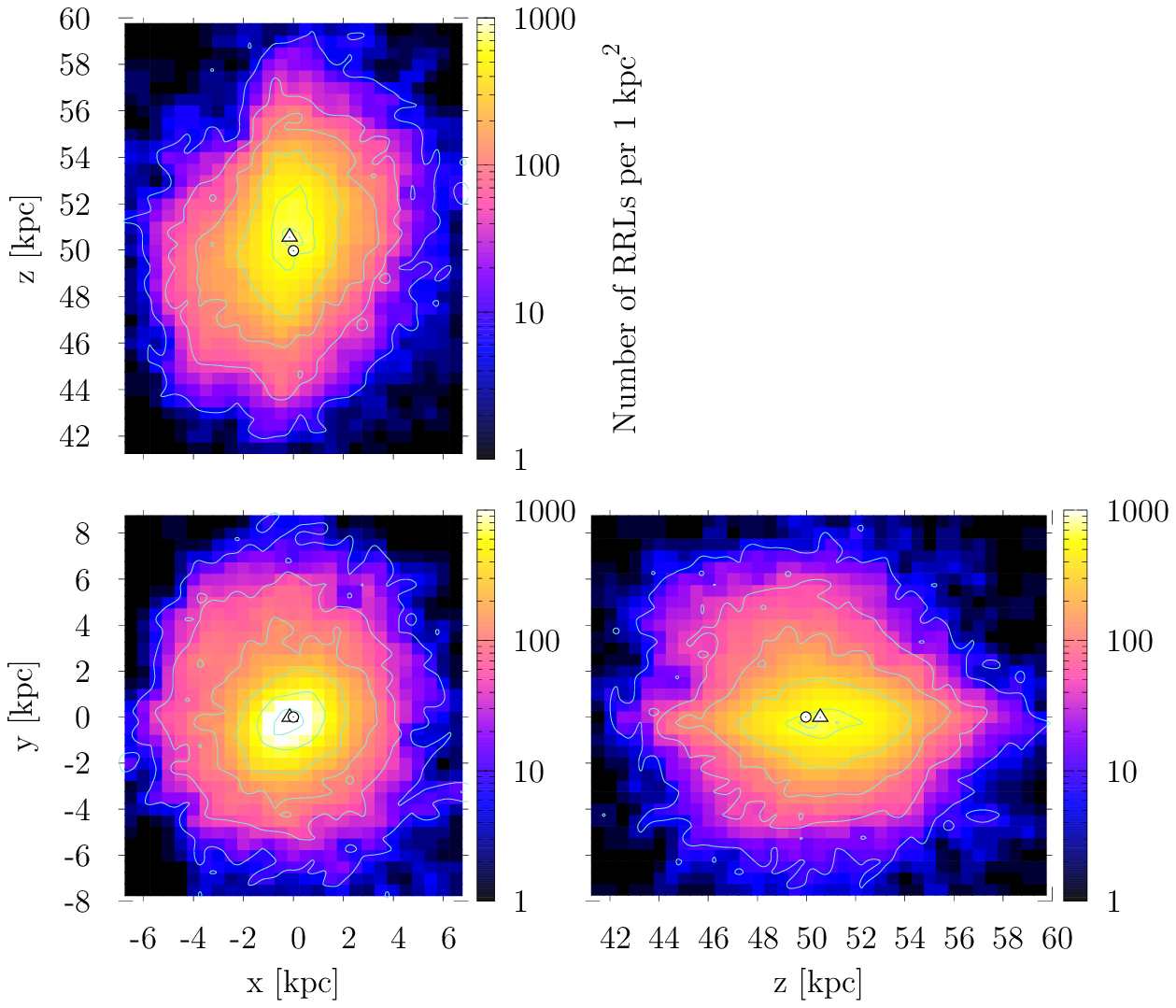} 
	\caption{Cartesian projections. {\it Left:} Classical Cepheids (Fig.~5b from \cite[Jacyszyn-Dobrzeniecka et al. 2016]{paperI}). {\it Right:} RR Lyrae stars (Fig.~7 from \cite[Jacyszyn-Dobrzeniecka et al. 2017]{paperII}).}
   \label{lmc}
\end{center}
\end{figure}

\section{Classical Pulsators in the SMC}

CCs in the SMC revealed a regular, non-planar structure that can be described as an ellipsoid elongated almost along the line of sight. Its longest axis is about five times longer than shorter axes. We have found two off-axis substructures in the SMC. These structures are not as prominent as those in the LMC. The northern SMC substructure is located closer to us and it is younger, while the southwestern SMC substructure is located farther and is older on average.
				
RRL stars in the SMC are distributed very regularly, even more than in the case of the LMC. We observe no additional substructures or irregularities. We also fitted triaxial ellipsoids to our sample. Virtually all of the resulting ellipsoids have the same shape (axes ratio). The outermost ellipsoids are more twisted towards the LMC.

\begin{figure}[htb]
\begin{center}
	\includegraphics[width=.49\textwidth]{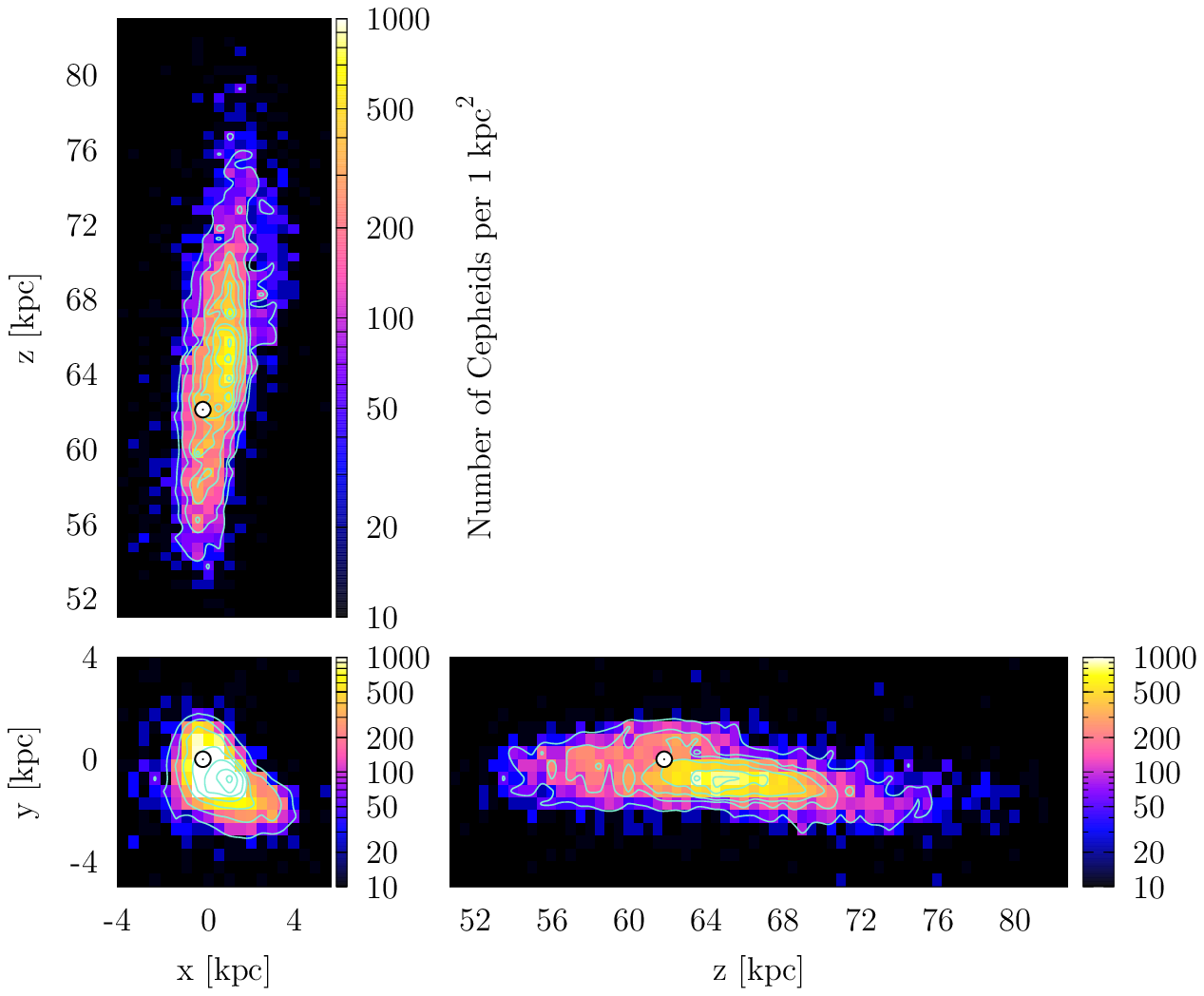} 
	\includegraphics[width=.5\textwidth]{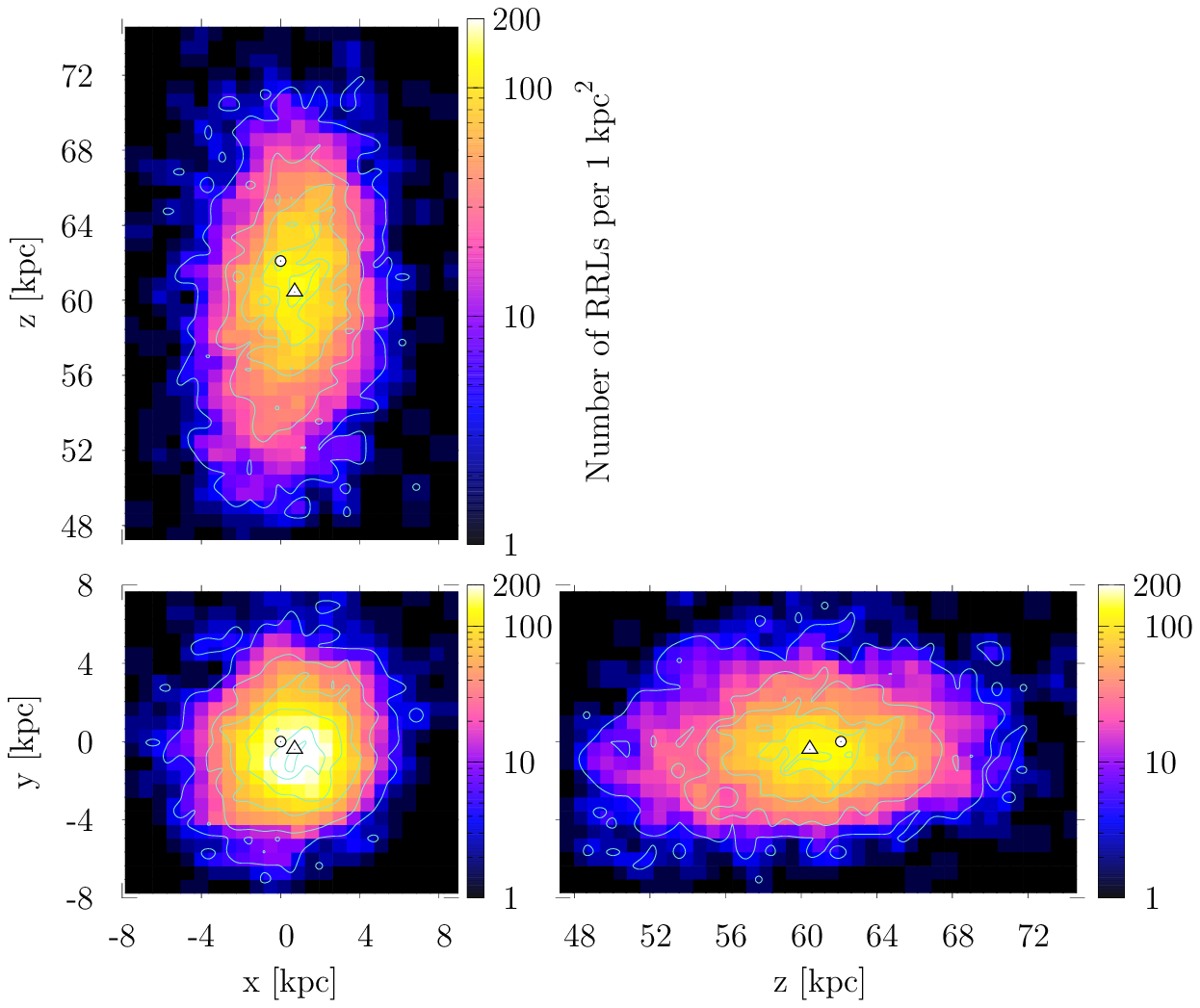} 
	\caption{Cartesian projections. {\it Left:} Classical Cepheids (Fig.~13b from \cite[Jacyszyn-Dobrzeniecka et al. 2016]{paperI}). {\it Right:} RR Lyrae stars (Fig.~12 from \cite[Jacyszyn-Dobrzeniecka et al. 2017]{paperII}).}
   \label{smc}
\end{center}
\end{figure}

\acknowledgements{A.M.J.-D. is supported by the Polish Ministry of Science and Higher Education under ``Diamond Grant'' No. 0148/DIA/2014/43.}

\end{document}